\documentclass[12pt]{iopart}

\usepackage{iopams}  
\usepackage{graphicx}
\usepackage{upgreek}
\usepackage[english]{babel}
\usepackage[T1]{fontenc}
\usepackage{setstack}
\usepackage[squaren,Gray]{SIunits} 
\newcommand{\tens}[1]{\overset{\leftrightarrow}{#1}}
\usepackage{epstopdf}

\begin{document}

\title{On three-dimensional dilational elastic metamaterials}

\author{ Tiemo B\"uckmann$^1$, Robert Schittny$^1$, Michael Thiel$^{1,2}$, Muamer Kadic$^1$, Graeme W. Milton$^{3}$,and Martin Wegener$^{1,4}$}

\address{$^1$Institute of Applied Physics, Karlsruhe Institute of Technology (KIT), 76128  Karlsruhe, Germany}
\address{$^2$Nanoscribe GmbH, Hermann-von-Helmholtz-Platz 1, 76344 Eggenstein-Leopoldshafen,  Germany}
\address{$^3$Department of Mathematics, University of Utah, Salt Lake City, Utah 84112, USA}
\address{$^4$Institute of Nanotechnology, Karlsruhe Institute of Technology (KIT), Hermann-von-Helmholtz-Platz 1,
76344 Eggenstein-Leopoldshafen, Germany}
\ead{Tiemo.Bueckmann@kit.edu}

\begin{abstract}
Dilational materials are stable three-dimensional isotropic auxetics with an ultimate Poisson's ratio of $-1$. We design, evaluate, fabricate, and characterize crystalline metamaterials approaching this ideal. To reveal all modes, we calculate the phonon band structures. On this basis, using cubic symmetry, we can unambiguously retrieve all different non-zero elements of the rank-4 effective metamaterial elasticity tensor, from which all effective elastic metamaterial properties follow. While the elastic properties and the phase velocity remain anisotropic, the effective Poisson's ratio indeed becomes isotropic and approaches $-1$ in the limit of small internal connections. This finding is also supported by independent static continuum-mechanics calculations. In static experiments on macroscopic polymer structures fabricated by three-dimensional printing, we measure Poisson's ratios as low as $-0.8$ in good agreement with theory. Microscopic samples are also presented.
\end{abstract}

\pacs{43.40.+s, 43.35.+d, 62.20.-x, 63.20.-e, 63.20.D-}
\noindent{\it Keywords}: mechanical metamaterials, acoustics, auxetics, dilational metamaterials
\maketitle

\tableofcontents

\section{Introduction}
Auxetic materials are rather special and unusual elastic solids. Their Poisson's ratio $\nu$ is negative, which means that it is easy to change their volume while fixing their shape, but it is hard to change their shape while fixing their volume \cite{Feynman2011}. This behavior is just opposite to that of an ideal liquid \cite{Born1939} and to that of an ideal pentamode metamaterial \cite{Kadic2012}. In general, auxetic materials can be anisotropic, in which case the Poisson's ratio turns into a Poisson's matrix \cite{Li1976, Rand2004}. There are no fundamental bounds for the values of the elements of the general Poisson's matrix \cite{Ting2005}. In sharp contrast, there are established bounds for stable elastic isotropic media. Here, the Poisson's ratio is connected to the ratio of bulk modulus $B$ (the inverse of the compressibility) and shear modulus $G$ via \cite{Rand2004}
\begin{equation}
\frac{B}{G}=\frac{1}{3} \frac{\nu+1}{0.5-\nu}.
\label{B_G}
\end{equation}
For a stable elastic solid, neither bulk nor shear modulus can be negative. For example, exerting a hydrostatic pressure onto a material with $B<0$ would lead to an expansion, further increasing the pressure, further increasing the volume, etc. This non-negativity together with equation (\ref{B_G}) immediately translates into the well-known interval of possible Poisson's ratios of $\nu \in [-1,0.5]$. 
Effectively isotropic auxetic materials with $\nu<0 $ composed of disordered polymer- or metal-based foams have extensively been studied in the literature, for a recent review see Ref.\,\cite{Greaves2011}. It is not clear though how one would systematically approach, along these lines, the ultimate limit of $\nu=-1 $. Such ultimate extreme auxetic materials are called ``dilational'' because they support strictly no other modes than dilations. Intuitively, for example, if one exerts a force onto a statue of liberty made of a dilational material at any point and along any direction, one can change its volume, but it will always maintain exactly the shape of the statue of liberty. Obviously, this behavior is very much different from that of a regular elastic solid. As an impact would be distributed throughout the entire elastic structure, dilational materials can, for example, be used as shock absorbers \cite{Alderson1999, Miller2009}. 
Early three-dimensional auxetic metamaterials with anisotropic Poisson's ratio have recently been presented \cite{Bueckmann2012}. It is again not clear though how this approach \cite{Bueckmann2012} could be brought towards an isotropic behavior with $\nu=-1$.  In the literature, several conceptual models for dilational metamaterials have been discussed \cite{Milton2013,Milton1992,Lakes2007,Prall1997,Mitschke2011,Grima2000}. These, however, contain elements like ``perfect joints'' and ``rigid rods'' that still need to be translated to a three-dimensional continuous microstructure composed of one constituent material (and vacuum in the voids) that can be fabricated with current technology. 
In this paper, inspired by the two-dimensional conceptual model of Ref.\,\cite{Milton2013}, we introduce a three-dimensional blueprint for such a dilational material. Several questions arise. Does this microstructure support unwanted easy modes other than the wanted dilations? Can this microstructure at all be described by a simple elasticity tensor and a constant mass density? For so-called Cosserat materials or for materials with anisotropic mass-density tensors \cite{Cosserat1909, Torrent2008, Milton2006, Schoenberg1983}, the answer would be negative. Our blueprint contains small internal connections mimicking the mentioned ``ideal joints''. How small do these connections have to be? The blueprint uses a simple-cubic translational lattice. Do we really get an isotropic Poisson's ratio? In general cubic elastic solids, the answer would be negative. To address all of these questions, we start by presenting calculated phonon band structures for our blueprint. Next, we compare these with static continuum-mechanics calculations. These can then directly be compared with our static experiments on macroscopic polymer-based metamaterials made by three-dimensional printing. Finally, we show that also microscopic versions can be fabricated by recent advances in galvo-scanner dip-in direct-laser-writing optical lithography.
A related but different idea has recently been demonstrated by a rubbery chartreuse ball with 24 carefully spaced round dimples \cite{Krieger2012}. These buckliballs can also be arranged into bucklicrystals \cite{Babaee2013}.

\section{The blueprint}
Our three-dimensional blueprint depicted in Fig.\,1 is based on a recently published two-dimensional conceptual model \cite{Milton2013}. This model contains ``ideal joints'' and ``rigid bars''. In our blueprint, the ideal joints are implemented by small connections between the square and the triangular elements. Upon pushing from one side, for example from the top, the inner squares rotate and the triangular outer connection elements get pulled inwards. Thus, ideally, the structure contracts laterally by the same amount as it contracts vertically. The Poisson's ratio would thus be $\nu=-1$. We will have to investigate though to what extend we approach this ideal for a finite connection size $d$ compared to the cubic lattice constant $a$. Furthermore, the very thin rigid bars in \cite{Milton2013} have been eliminated in our blueprint because they cannot be implemented using a single constituent material. As a result, it is not clear whether unwanted easy modes of deformation might occur. Indeed, in preliminary simulations, we have found that using only one sense of rotation of the squares, the squares do not only rotate around their center, but rather also translate. To eliminate this unwanted easy mode, we use a three-dimensional checkerboard arrangement with the discussed motif alternating with its mirror image. The small cubes with a side length identical to the thickness of all squares and triangles are not necessary for the function of the metamaterial. They are, however, crucial as markers in our measurements of the Poisson's ratio (see below). They are hence considered in all our band structure and static calculations to allow for direct comparison.
\begin{figure}
\includegraphics[scale=1]{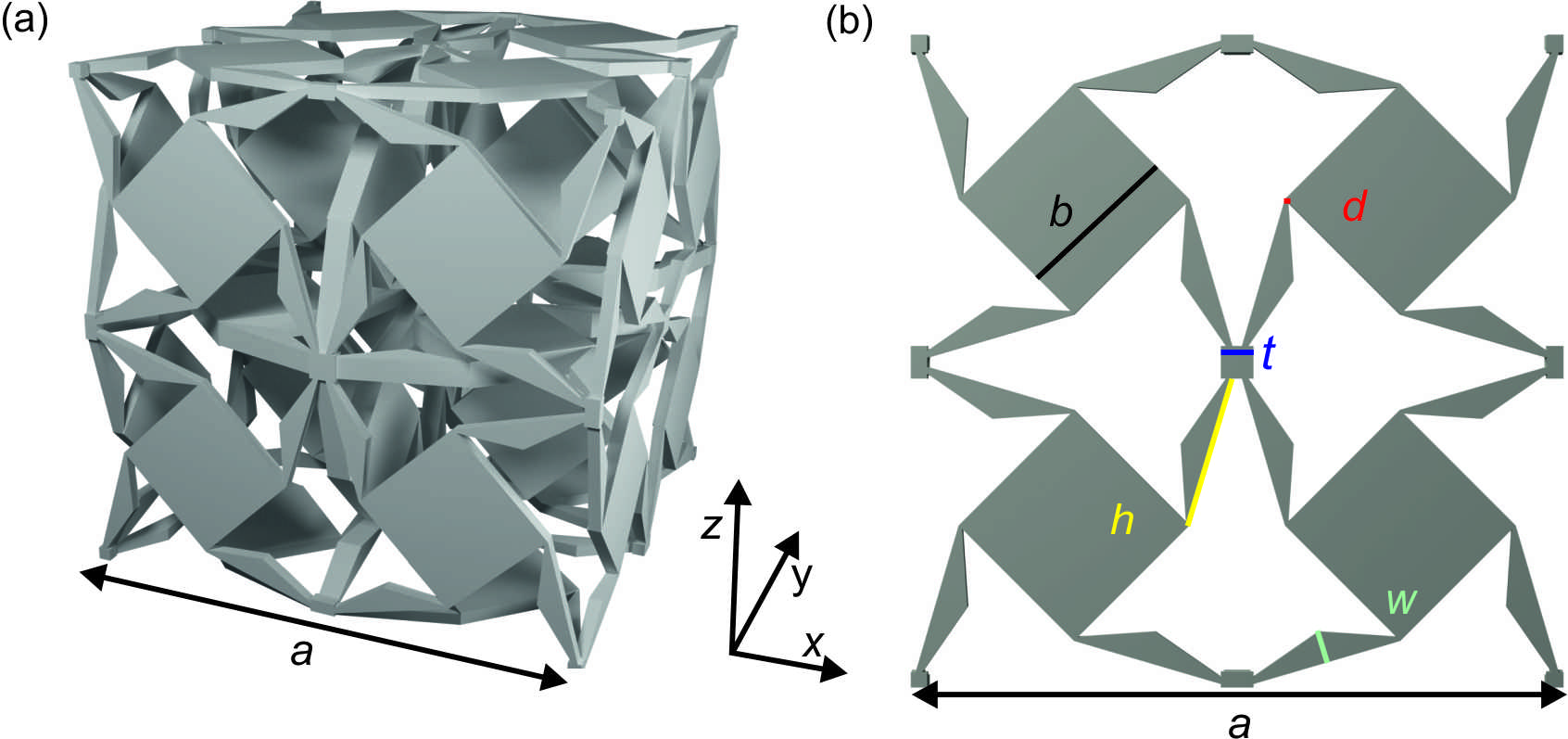}
\caption{(a) Illustration of our blueprint for a three-dimensional dilational metamaterial. A single unit cell is shown. This cell is composed of a checkerboard arrangement of a cubic motif and its mirror image. (b) One face of the structure with geometrical parameters indicated. The cubic lattice constant is $a$. The other parameters are: block size $b/a=0.25$, width of the holding element $w/a=0.048$, layer thickness $t/a=0.05$, holder length $h/a=0.235$, and connection size $d/a=0.5 \%$.}
\label{Fig1}
\end{figure}
\section{Phonon band structures}
The phonon band structure reveals all modes of the elastic metamaterial, possibly including unwanted easy modes other than dilations (see above). The long-wavelength limit of the band structure can be the starting point for a description in terms of effective elastic metamaterial parameters (see next section).
In our numerical band-structure calculations for the dilational metamaterial structure in Fig.\,1, we solve the usual elastodynamic equations \cite{acouMetaPhoCry} for the displacement vector 
$\vec{u}(\vec{r},t) $ containing the time-independent rank-4 elasticity tensor $\tens{C} \left( \vec{r}\right)$ and the scalar mass density $ \rho(\vec{r}) $, i.e., 
\begin{equation}
\vec{\nabla} \cdot(\tens{C} \vec{\nabla} \vec{u}  )-\rho  \frac{\partial^2 \vec{u}}{\partial t^2 }=0,
\label{eq2}
\end{equation}
by using a commercial software package (COMSOL Multiphysics, MUMPS solver). We impose Bloch-periodic boundary conditions onto the primitive cubic real-space cell shown in Fig.\,\ref{Fig1}. We have carefully checked that all results presented in this paper are converged. Typically, convergence is achieved by using several ten thousand tetrahedra in one primitive real-space cell. We choose an isotropic polymer as constituent material with Young's modulus $1\, \rm{GPa}$, Poisson's ratio $0.4$, and mass density $1200\,\rm{kg/m^3}$. These values are chosen according to the below experiments. Due to the scalability of the elastodynamic equations, our results can easily be scaled to isotropic constituent materials with any different Young's modulus and density. The Poisson's ratio of the constituent material influences the results only to a very minor degree. The voids in the polymer are assumed to be vacuum. The lattice constant is chosen as $a=4.8\,\rm{cm}$, according to our below experiments on macroscopic polymer structures. However, the results can again easily be scaled to any other value of $a$. We represent the results in a simple-cubic Brillouin zone corresponding to the underlying simple-cubic translational lattice. 
\begin{figure}
\begin{center}

\includegraphics[scale=1]{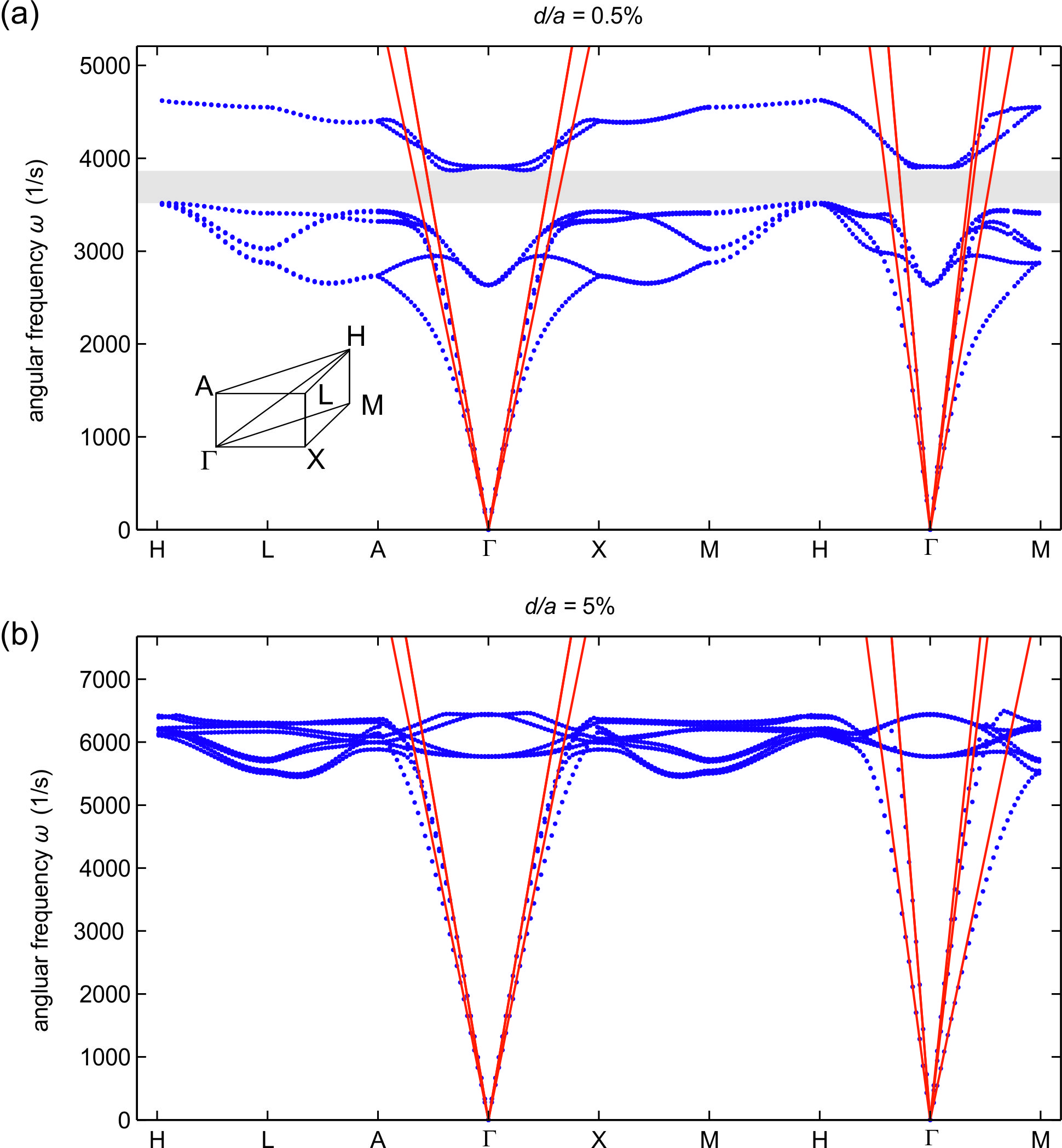}
\caption{ Calculated phonon band structures (blue dots) of dilational metamaterials, i.e., angular frequency $\omega$ versus wave vector for the usual tour through the simple-cubic Brillouin zone (see inset in (a)). The red straight lines are fits assuming a homogeneous cubic-symmetry effective medium. (a) connection size $d/a=0.5 \% $ (b) $d/a=5\%$; $a= 4.8\,\rm cm$. The grey area in (a) highlights a complete three-dimensional elastic band gap.}
\label{Fig2}

\end{center}
\end{figure}
Examples of calculated band structures are depicted in Fig.\,\ref{Fig2} for two different values of the ratio $d/a$. Shown are the six lowest eigenmodes. It becomes immediately clear that the slope of the low-frequency or long-wavelength acoustic modes is not the same for all directions, it is anisotropic. Along the $\Gamma\rm X $-direction (i.e., along the principal cubic axes), the velocity of the longitudinally polarized mode is smaller than that of the transversely polarized modes (see Fig.\,\ref{Fig3}). In isotropic elastic media, the opposite holds true. As to be expected, the phase velocities are larger for larger $d/a$. 
We note in passing that the band structure in Fig.\,\ref{Fig2} for $d/a=0.5 \% $ exhibits a complete three-dimensional elastic (i.e., not only acoustic) band gap between normalized frequencies of $3.24\,\rm{kHz}$ and $3.55\,\rm{kHz}$. This region with zero bulk phonon density of states corresponds to a gap-to-midgap ratio of $9 \%$. This complements other possibilities reported in the literature \cite{acouMetaPhoCry, Wang2012}. 
To further emphasize the anisotropy, we also plot the phase velocity in polar diagrams in Fig.\,\ref{Fig3}. Constant phase velocity would lead to circles in the two-dimensional cuts depicted. Clearly, the curves shown in Fig.\,\ref{Fig3} (a) and (b) are not circular at all, neither for the small nor for the large value of $d/a$. They do show four-fold symmetry though. This statement is not trivial, because the three orthogonal axes are not strictly equivalent in terms of the geometrical structure. This can be seen when comparing a view on the unit cell shown in Fig.\ref{Fig1} onto the $xy$- and the $xz$-planes. Nevertheless, the band structures show that wave propagation is equivalent for the three cubic axes. 
\begin{figure}
\includegraphics[scale=1]{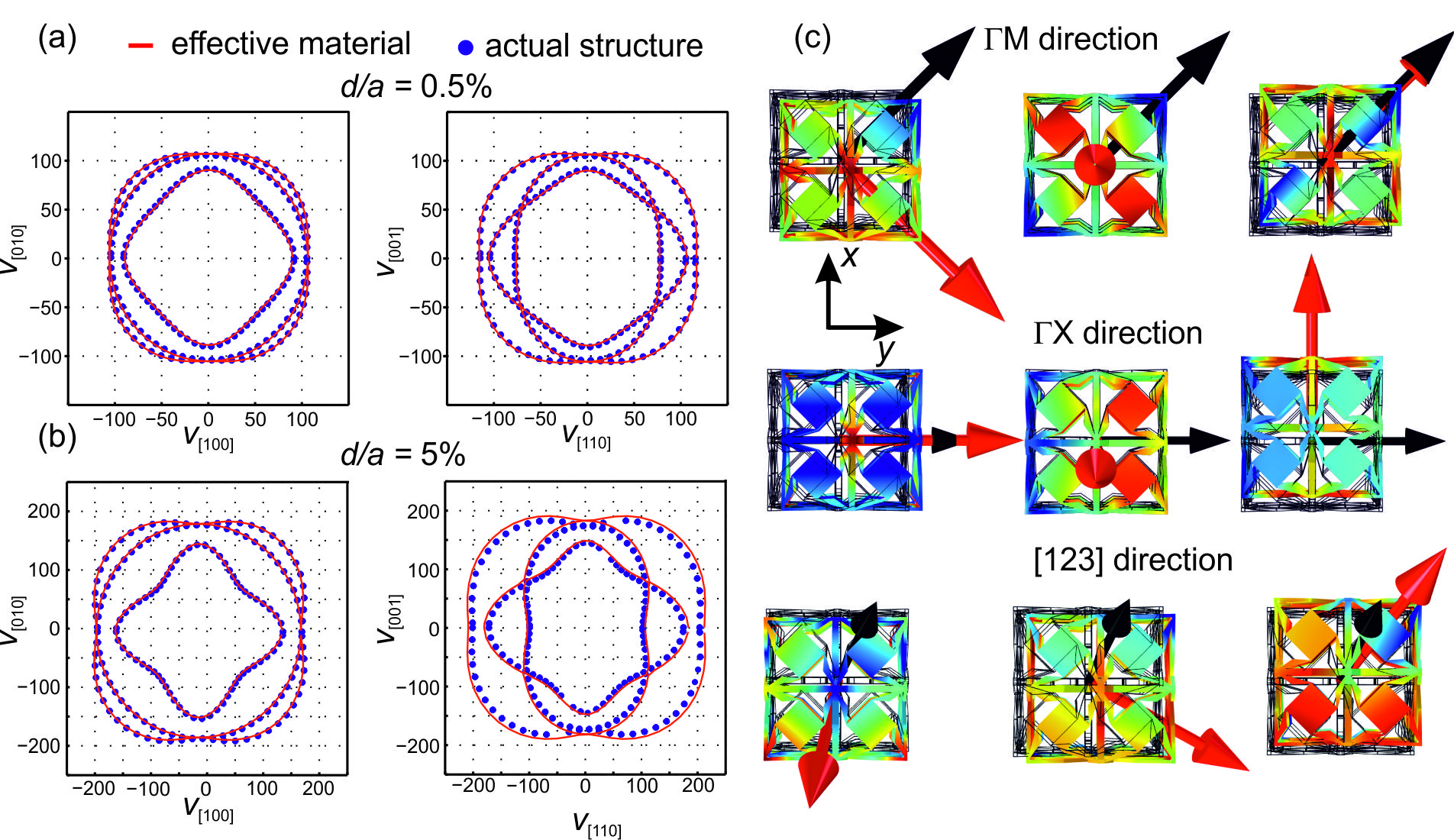}
\caption{(a) and (b) are polar representations of the phase velocity at a wave number of $k=0.01/a$, i.e., the phase velocity in a particular direction is given by the radial length. The cut on the left is for the $xy$ plane, the cut on the right for a plane spanned by the [111] and the [110] directions. The cut on the left shows the fourfold rotational symmetry expected for a cubic structure. The connection size is (a) $d/a=0.5\%$ and (b) $d/a=5\%$. All other geometrical parameters are as quoted in Fig.\,1. The blue dots are derived from the phonon band structure, the red curves are the result of an effective-parameter description of a cubic-symmetry medium. (c) Selected eigenmodes for a fixed wave number of $k=0.2/a$ and for $d/a=0.5\%$. The corresponding eigenfrequencies increase from the left to the right. The black arrows indicate the direction of the wave vector $\vec{k}$, the red arrows the directions of the displacement vector $\vec{u}$. Shown are the $\Gamma \rm X$ direction identical to the principal cubic axes, the $\Gamma \rm M$ direction parallel to the cubic face diagonals, as well as for an oblique direction. For the latter, the modes are no longer purely transversely or longitudinally polarized.}
\label{Fig3}
\end{figure}

The anisotropic behavior of the phase velocity is connected to rather complex underlying eigenmodes that are illustrated by the examples shown in Fig.\,\ref{Fig3} (c). For waves propagating along the principal cubic axes or along the face diagonals, we find pure longitudinal or pure transverse polarization. For arbitrary oblique propagation directions with respect to the principal cubic axes, the eigenmodes are complicated mixtures of transverse and longitudinal polarization. In contrast, in an ideal isotropic elastic medium, the polarizations would be purely transverse or longitudinal.

\section{Retrieval of the elasticity tensor}
The phonon band structures presented in the previous section have shown pronounced anisotropies resulting from the cubic symmetry of the underlying translational lattice. In this light, one might suspect that the Poisson's ratio is anisotropic as well, whereas we aim at an isotropic Poisson's ratio. 
We thus derive a Poisson's ratio from the phonon band structure. To do so, we compare the band structures with the expectation from continuum mechanics of an effective cubic-symmetry medium. For crystals obeying simple-cubic symmetry \cite{Paszkiewicz2008}, the rank-4 elasticity tensor $\tens{C}$ has the three different non-zero elements $C_{11}= C_{22}= C_{33}= C_{1111} = C_{2222} = C_{3333}, C_{12}= C_{13}=C_{23}= C_{1122}= C_{2211}= C_{1133}= C_{3311}= C_{2233}= C_{3322},\text{and } C_{44}= C_{55}= C_{66}= C_{2323}=C_{3232}=C_{2332}= C_{3223}= C_{1313}= C_{3131}= C_{1331}= C_{3113}= C_{1212}= C_{2121}= C_{1221}= C_{2112}$. Here, the elements with only two indices refer to Voigt notation \cite{Voigt1910}. All other elements are zero. Furthermore, we assume a constant scalar effective metamaterial mass density $\rho $, which is simply given by the volume filling fraction $f$ of the constituent material times its own bulk mass density (see Fig.\,1). For $d/a=0.5 \%$ ($d/a=5 \% $) we get $f=10.4\%$ ($f=11.2\% $). On this basis, we can now calculate the phonon band structure in the long-wavelength limit. To do so, one needs to connect the phase velocities with the elements of $\tens{C}$ and with $\rho$. Here, it is convenient to inspect the $\Gamma \rm M$ or [110] direction with three different phase velocities $v$ and three orthogonal eigenmodes that are either purely longitudinally (L) or purely transversely (T) polarized (see above). The latter either lie in the $xy$ plane or along the $z$-direction. Following \cite{Tsang1983}, the connections are given by
\begin{equation}
	C_{44}= \rho \,(v_{110}^{{\rm T},z})^2,
	\end{equation}
	\begin{equation}
C_{12}= \rho\,(v_{110}^{\rm L})^2  - C_{44} - \rho \,(v_{110}^{{\rm T},xy})^2,
	\end{equation}
	\begin{equation}
C_{11}= 2\, \rho\,(v_{110}^{{\rm T},xy})^2 +C_{12}.
\end{equation}
The three different elements of $\tens{C}$ can immediately be computed from the three different phase velocities. One has to make sure though that the polarizations of the corresponding eigenmodes are the same as for the numerical band structure calculations. We have checked this aspect (not depicted). Furthermore, one needs to make sure that the elastic behavior is described for all other propagation directions as well. To this end, we compare in Figs.\,\ref{Fig2} and \ref{Fig3} the results from the phonon band structure (blue dots) with those of the effective-medium description (red lines). Obviously, we obtain excellent overall agreement for all conditions in the long-wavelength limit. This means that a description of the elastic metamterial in terms of an elasticity tensor $\tens{C}$ for a cubic-symmetry effective medium is adequate. As discussed in our introduction (see Cosserat and anisotropic-mass-density metamaterials), this finding itself is not trivial. 
Having derived all non-zero elements of the effective metamaterial elasticity tensor, we can now apply established analysis to extract the Young's modulus $E$, the shear modulus $G$, the bulk modulus $B$ \cite{Bowers2009, Hill1952}, and the Poisson's ratio (or Poisson's matrix) \cite{Wojciechowski2005}. We have
\begin{equation}
E=\frac{C_{11}^2+C_{12}C_{11}-2C_{12}^2}{C_{11}+C_{12}}, 
	\end{equation}
	\begin{equation}
G=C_{44},
	\end{equation}
	\begin{equation}
B=  \frac{C_{11}+2C_{12}}{3}.
	\end{equation}
Examples are given in Table 1.
\begin{table}[h]
\caption{Examples of retrieved effective parameters. The three non-equivalent non-zero elements of the elasticity tensor $C_{11}$, $C_{12}$, $C_{44}\,=\,G$, the Young's modulus $E$, and the bulk modulus $B$ are given for selected values of $d/a$.\\}

\begin{tabular}{c|c|c|c|c|c}
  $d/a$ (\%) & $C_{11}$ (MPa) &  $C_{12}$ (MPa)&  $C_{44}\,=\,G$ (MPa) & $E$ (MPa)& $B$ (MPa)  \\ \hline
  $5$ &$2.33$ &$0.0051$ &$3.5 $& $2.33$ &$ 0.78$\\
  $4 $& $2.33$&$-0.16 $& $3.78$& $2.31$& $0.67$\\
  $3$ & $2$&$-0.35$ &$3.36$ & $1.85$& $0.43$\\
  $2.5$ & $1.3$& $-0.52$&$2.1$ & $6.07$&$ 0.0867$\\
  $0.75$ &$1.14$ &$-0.507$ &$1.68$ &$ 0.33$& $0.042$\\
  $0.5 $& $0.97$& $-0.445$&$1.35$ & $0.21$& $0.026$\\
  $0.25$ &$0.85$ & $-0.406$&$1.06$ & $0.12$&$0.014$ \\
\end{tabular}

\end{table}

\begin{figure}
\begin{center}
\includegraphics{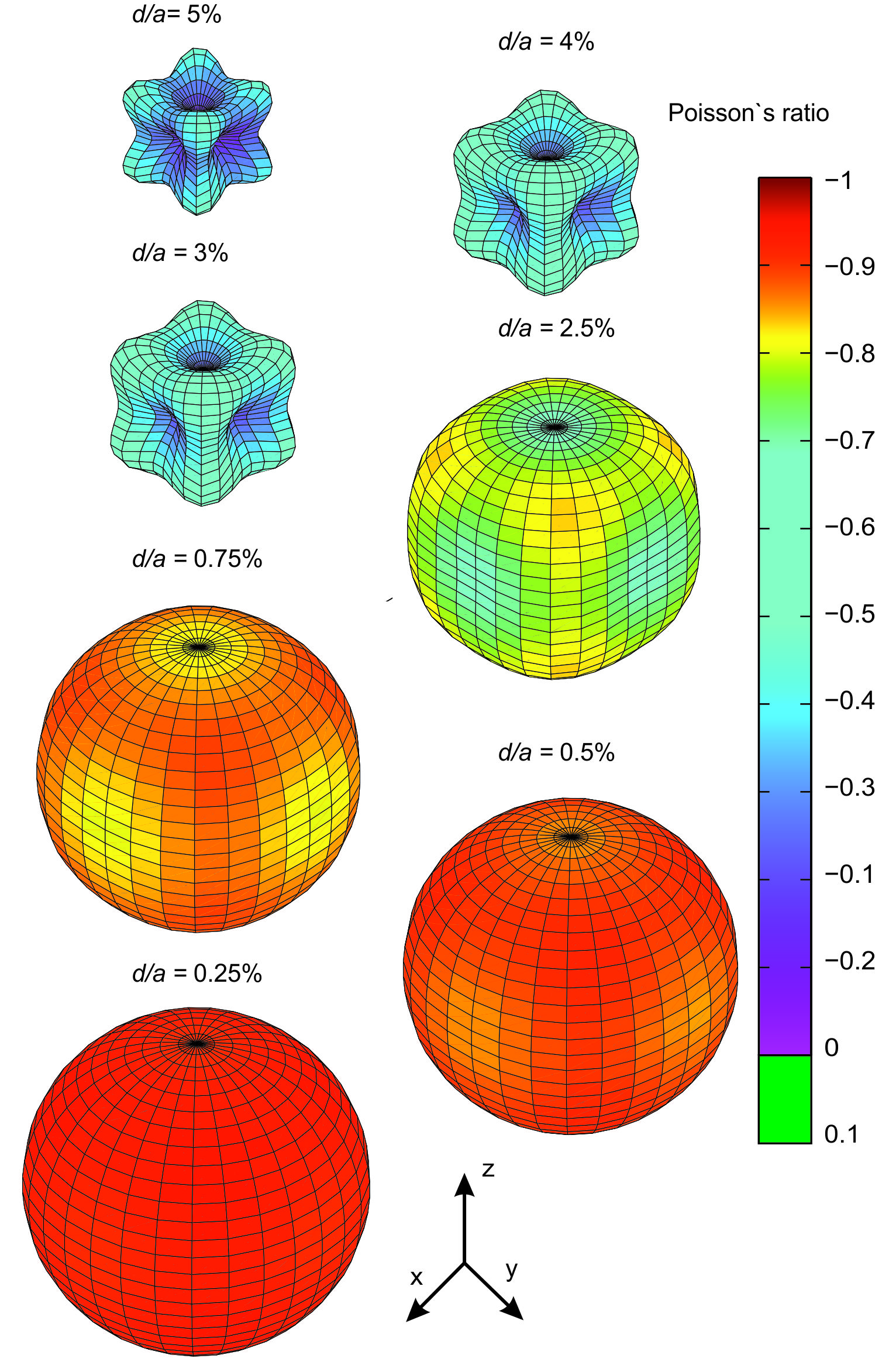}
\caption{Three-dimensional polar diagram of the effective metamaterial Poisson's ratio $\nu$, i.e., the length of the vector from the origin to the surface is proportional to modulus of the Poisson's ratio. The Poisson's ratio including its sign is also indicated by the false-color scale. As the $d/a$ ratio decreases, $\nu$ becomes more negative and more isotropic, eventually approaching the ultimate limit of $-1$ for an isotropic elastic material. These results are derived from the band structures like exemplified in Fig.\,\ref{Fig2} and for the other geometrical parameters as in Fig.\,\ref{Fig1}.}
\label{Fig4}
\end{center}
\end{figure}

\begin{figure}
\includegraphics[scale=1]{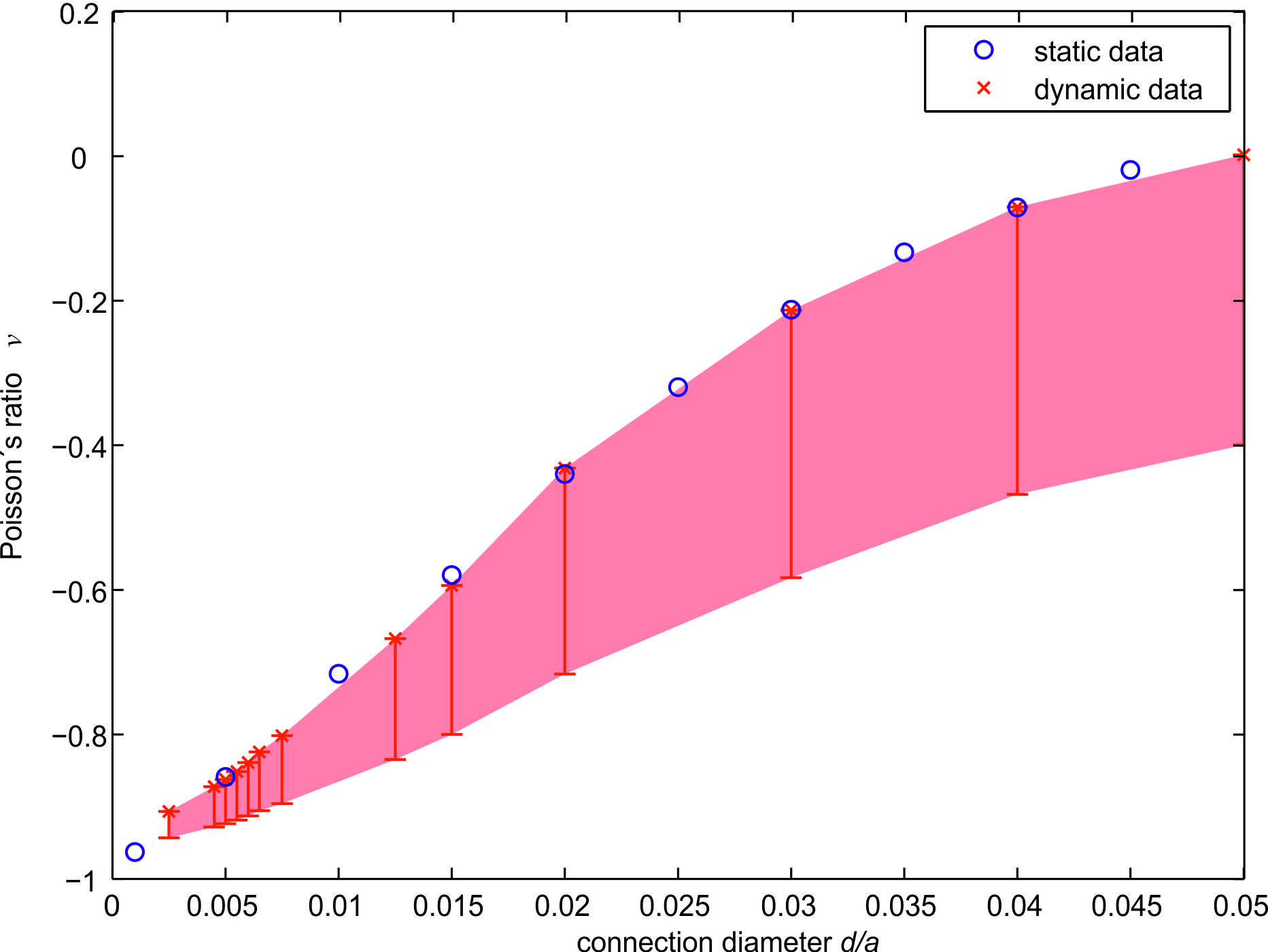}
\caption{Minimum and maximum effective metamaterial Poisson's ratio (see Fig.\,\ref{Fig4}) versus $d/a$ ratio. The red symbols are the minima and maxima derived from the phonon band structures, the blue dots are obtained from static continuum-mechanics calculations for pushing along one of the principal cubic axes.}
\label{Fig5}
\end{figure}

The Poisson's ratio for pushing along the principal cubic axes is given by \cite{Bowers2009, Hill1952}
\begin{equation}
\nu =\frac{C_{12}}{C_{11}+C_{12}}.
\end{equation}
However, the Poisson's ratio might still be different for arbitrary oblique pushing directions. Here, we use the more general expressions as given in Ref. \cite{Wojciechowski2005}, which are based on averaging along the directions normal to the pushing direction:
\begin{equation}
\nu (\phi ,\theta )=-\frac{Ar_{12}+B(r_{44}-2)}{16[C+D(2r_{12}+r_{44} )] }
\label{eq:nuang}
\end{equation}
Introducing the compliance tensor $\tens{S}=\tens{C}^{-1}$ (two indices refer to Voigt notation), the abbreviations in (\ref{eq:nuang}) are given by

\begin{eqnarray*}\label{eqn:nu_definitions}
  		r_{12}&=&\frac{S_{12}}{S_{11}},  \\
  		r_{44}&=&\frac{S_{44}}{S_{11}}, \label{subeqn-2:nu_definitions}\\
  		A&=&2\left[53+4 \rm{cos}{(2 \theta)}+7 \rm{cos}{(4 \theta)}+8 cos{(4\phi)} sin^4{(\theta)}\right],\label{subeqn-3:nu_definitions}\\ 
		B&=&-11+4\rm{ cos}{(2 \theta)}+7 cos{(4 \theta)}+8 cos{(4\phi)} sin^4{(\theta)}, \label{subeqn-4:nu_definitions}\\
		C&=&8 \rm cos^4 { (\theta)}+6 sin^4 { (\theta)}+2 cos{(4\phi)} sin^4{(\theta)}, \label{subeqn-5:nu_definitions}\\
   		D&=&2\left[ \rm sin^2 {( 2 \theta) }+ sin^4 { (\theta) } sin^2 {(2 \phi)}\right].\label{subeqn-6:nu_definitions}
\end{eqnarray*}
Here, $\phi$ and $\theta$ are the usual angles in spherical coordinates. 
The resulting direction dependence of the Poisson's ratio is visualized in two different ways in Fig.\,\ref{Fig4} and Fig.\,\ref{Fig5}. Fig.\,\ref{Fig4} is the generalized polar representation, i.e., the Poisson's ratio is proportional to the length of the vector from the origin to the depicted surface. The Poisson's ratio is also visualized by the false-color scale. For large connections, e.g., for $d/a=5\%$, the behavior is obviously far from isotropic. Furthermore, the effective metamaterial Poisson's ratio is far from $-1$. For a decreasing connection size $d/a$, the effective metamaterial Poisson's ratio $\nu$ becomes more negative and reaches a nearly isotropic behavior at $d/a=0.25\%$. This means that experiments need to realize values of $d/a$ below $1\%$ or better. Fig.\,\ref{Fig5} depicts the derived minimum and maximum values of $\nu$ versus $d/a$. For the smallest numerically accessible values of $d/a$, the Poisson's ratio comes close to $-1$. The data do not appear to extrapolate to exactly $-1$ though. We suspect that this aspect is due to the small cubes in our blueprint that we have introduced for experimental reasons (see above). We expect that the effective metamaterial Poisson's ratio would converge to $-1$ in the hypothetical limit of no cubes and $d/a\rightarrow 0$ -- which cannot be realized experimentally though.

\section{Static continuum-mechanics calculations}
\begin{figure}
\includegraphics[scale=1]{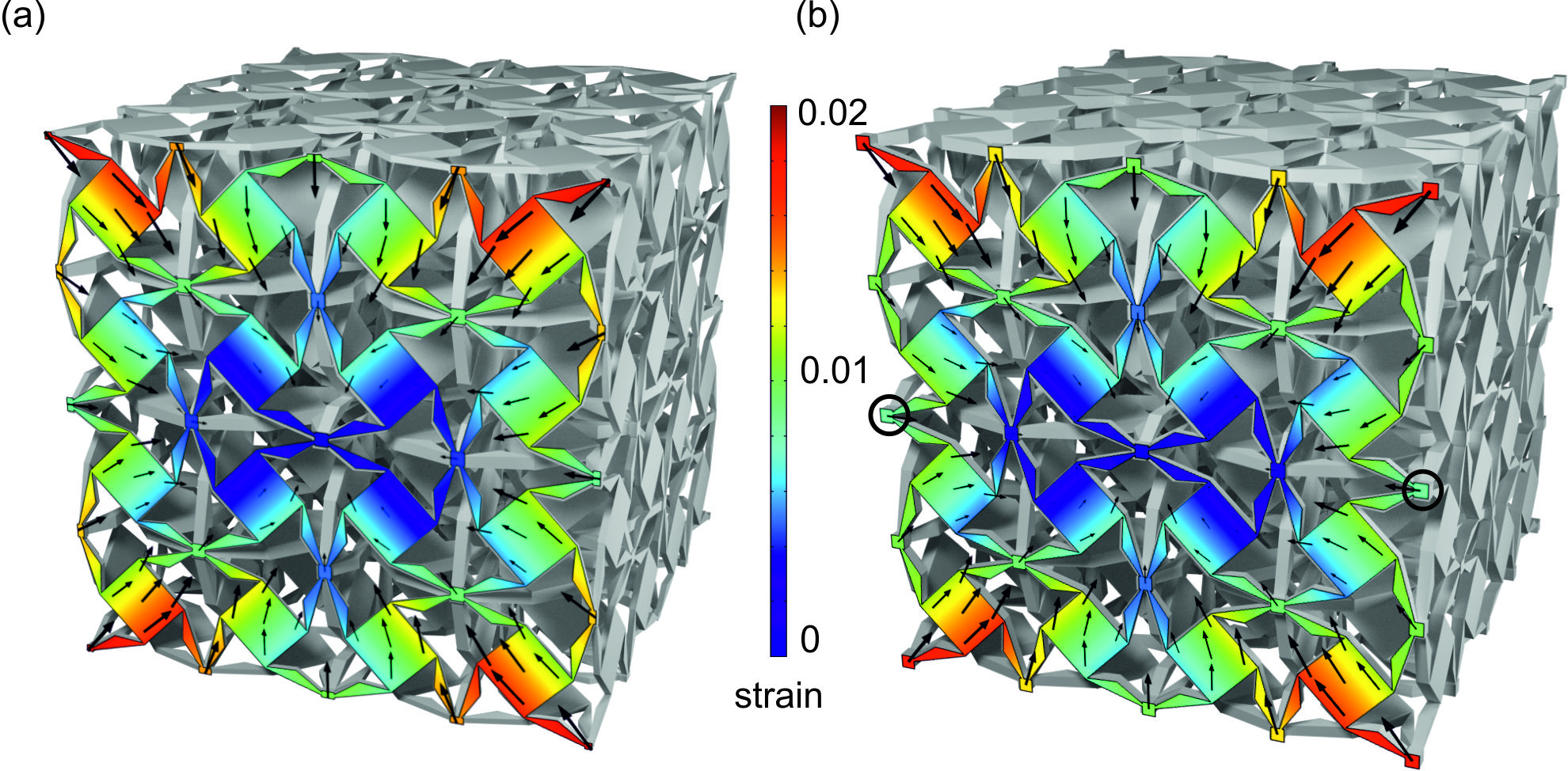}
\caption{(a) $2\times2\times2$ unit cells (compare Fig.\,\ref{Fig1}) out of an infinite crystal pushed upon along the $z$-direction. The resulting in-plane strain for an axial strain of 0.01 is depicted by the false-color scale projected onto the front surface of the cube as well as by the black arrows. (b) Same, but for a finite crystal with $2\times2\times2$ unit cells.}
\label{Fig6}
\end{figure}

Often, the Poisson's ratio is measured in (quasi-)static experiments. To derive the Poisson's ratio, one pushes along one direction, e.g., the $z$-direction, leading to a certain strain (or relative displacement) $\epsilon_{zz}$, observes or calculates the displacement along the orthogonal $x$-direction (or the orthogonal $y$-direction), hence the element of the strain tensor $\epsilon_{xx}$, and computes the Poisson's ratio according to its definition

\begin{equation}
\nu = -\frac{\epsilon_{xx}}{\epsilon_{zz}}.
\end{equation}
Furthermore, experiments are based on finite-size samples, also containing a finite number of unit cells only. For metamaterials, the number of unit cells may be rather small. We thus also investigate the question to what extend measurement artifacts are to be expected for accessible-size samples.
The numerical calculations to be presented in this section have been performed with COMSOL Multiphysics using the Structural Mechanics module. The structure's geometry is created using the CAD COMSOL Kernel. The mesh is created within COMSOL Multiphysics using the preset parameter values called ``normal'' meshing with settings(referring to a $1\times 1\times 1 \rm m^3$ geometry size): Maximum Element Size = 0.1, Minimum Element Size = 0.018, Maximum Element Growth Rate = 1.5, Resolution of Curvature = 0.6, Resolution of Narrow Regions = 0.5. For example, for a connection size of $ d/a=0.5 \% $, this leads to about 90000 tetrahedral elements. We use the MUMPS Solver. In convergence tests, we have verified that the derived effective metamaterial Poisson's ratios are accurate to within 0.01. All geometrical parameters and constituent material parameters are as given above.
To mimic a fictitious infinitely extended crystal, we assume that all unit cells behave the same way (analogous to zero wave vector in the previous section). For convenience, we choose our coordinate system such that the crystal center of mass is fixed. For pushing along one principal cubic axis these conditions can be implemented by imposing anti-symmetric boundary conditions onto a single cubic unit cell. This means that the normal component of the displacement vector on one surface of the unit cell is constant on this surface and equal to the negative of the normal component on the opposing surface. To investigate the linear regime, we choose strains along the pushing direction of $1\% $. 
The resulting behavior is illustrated in Fig.\,\ref{Fig6}(a). The length of the (black) arrows is exaggerated and indicates the local displacement vectors. The false-color scale shows the modulus of the local displacement vector. Note that the corners of the unit cell move nearly diagonally towards the center. The Poisson's ratio can immediately be computed from the components of these displacement vectors. For example, for $d/a=0.75\%$ corresponding to our below experiments, we obtain $\nu=-0.79$. We note in passing that the other points within the unit cell generally move in different directions than the corners. This has important implications for our below experiments in that we must not evaluate the movement of all points within the unit cell but rather only of the corners. To have a finite region for imaging and tracking, we have introduced the small cubes in Fig.\,\ref{Fig1}. The results of $\nu$ versus $d/a$ are shown in blue in Fig.\,\ref{Fig5}. Obviously, the agreement of these (static) values with those derived from the (dynamic) phonon band structures is good, providing further faith into the validity of our results. The most negative dynamic results tend to be more negative than the static ones. For example, for $d/a=0.75\%$, the minimum dynamic value is $\nu=-0.895$, the static one $\nu=-0.79$.
Static calculations have also been performed for a finite metamaterial sample containing $2\times2\times2$ unit cells (see Fig.\,\ref{Fig1}) as shown in Fig.\,\ref{Fig6}(b) to directly compare these with experiments. Here, we assume sliding boundary conditions parallel to the pushing stamp interfaces. Good agreement together with calculations for infinite crystals as above will allow us to extract Poisson's ratios under these extreme conditions. Fig.\,\ref{Fig6}(b) is an example for a finite crystal composed of $2\times 2\times 2$ unit cells. To compare to the experiments to be discussed below, one can measure the lateral strain of the left and right outer corners in the middle of the horizontal direction (see small circles) and divide these by the relative axial shift of the stamps. We obtain a Poisson's ratio for the finite crystal of $-0.76$, which is not too far from the one for the fictitious infinitely extended crystal of $\nu=-0.79$ in panel (a) of the same figure. These parameters have been chosen to match those of the experiments to be discussed next.

\section{Macroscopic dilational metamaterials}
\begin{figure}

\includegraphics[scale=1]{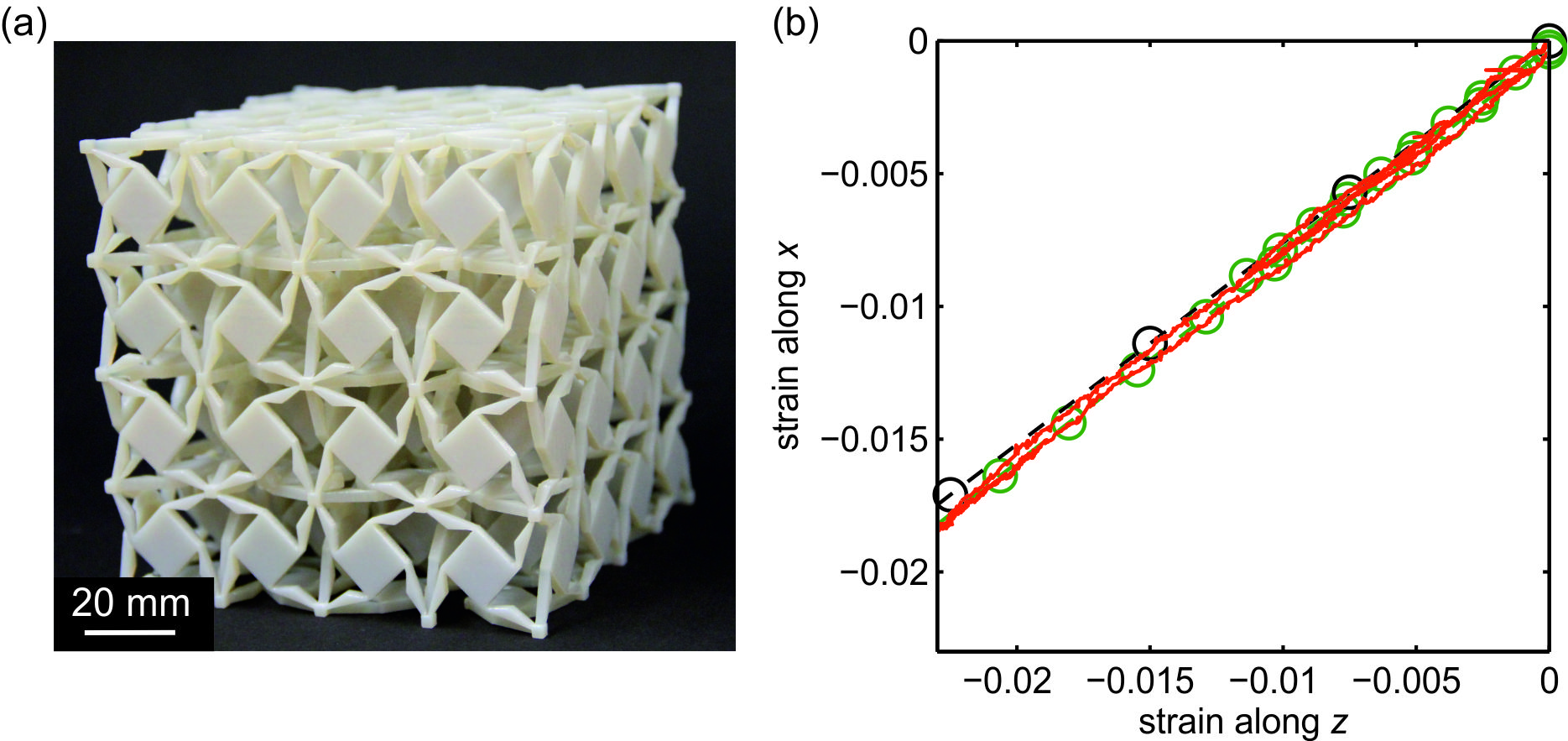}
\caption{(a) Photograph of a macroscopic polymer-based finite crystal with $2\times2\times2$ unit cells fabricated by 3D printing, following the blueprint and the parameters given in Fig.\,\ref{Fig1}. (b) Measured lateral versus axial strain (solid red curve) as obtained from an image correlation approach upon pushing along the vertical $z$-direction. The green circles on straight lines correspond to a Poisson's ratio of $-0.76$ and $-0.77$, respectively, the black circles to numerical calculations for $d/a=0.75\%$. Extrapolation to an infinite three-dimensional crystal (see previous section) delivers a Poisson's ratio of $\nu=-0.79$.}
\label{Fig7}
\end{figure}

We have fabricated macroscopic (this section) and microscopic (next section) versions of the blueprint shown in Fig.\,\ref{Fig1}. The macroscopic samples are fabricated with the printer ``Objet30'' sold by former Objet, now Stratasys, USA. For the metamaterial, we have used the basic polymer ink ``FullCure850 VeroGray''. During the fabrication, however, one also needs a support material. The default is a mixture of ``FullCure850 VeroGray'' and ``FullCure705 Support'' that we have not been able to remove from the composite. Thus, we have chosen exclusively ``FullCure705 Support'', which can be etched out in a bath of NaOH base after hand cleaning. The structure files are exported in the STL file format directly from the geometry used in the COMSOL Multiphysics calculations and are imported into the Objet printer. The printing is done automatically in a standard process. An example of a fabricated macroscopic metamaterial structure is shown in Fig.\,\ref{Fig7}(a). The geometrical parameters are as defined in Fig.\,\ref{Fig1} with $d/a=0.75\%$ and $a=4.8\,\rm cm$. The constituent polymer material has a measured Young's modulus of about $E\approx 1.5\,\rm GPa$. 
The measurement setup for the macroscopic samples consists of two metallic stamps and a linear stage containing a force cell. The sliding boundary conditions (see previous section) are achieved by placing the watered sample between the stamps.  We have alternatively attempted to implement fixed boundary conditions by gluing the sample to the stamp by double-sided tape. This has led to the same lateral displacements of the sample indicating very strong forces parallel to the stamps. This suggests to us that assuming sliding boundary conditions is adequate. We gradually push onto one stamp by moving the linear stage while fixing the other stamp and recording the images of one of the sample surfaces. These images are taken with a Canon EOS 550D camera in Full HD ($1920 \times 1080$ pixels) resolution and 24 frames per second. 
The objective lens (Tamron SP 70-300mm f/4-5,6 Di VC USD) is located at a distance of approximately $1.5\,\rm m$ to the sample. We have checked that image distortions (e.g., barrel-type aberrations) are sufficiently small to not influence our experiments. The displacements of the unit cells' corners are tracked using an autocorrelation approach used previously \cite{Bueckmann2012}. Multiple measurements with increasing maximum strain for a crystal composed of $2 \times 2 \times 2$ unit cells (see Fig.\,\ref{Fig7}(a)) are depicted in Fig.\,\ref{Fig7}(b). The graph shows the strain along the horizontal $x$-direction versus the strain along the axial pushing direction ($z$). The solid red curve corresponds to two measurement cycles, i.e., the sample is pushed, released and pushed and released again. Clearly, the four parts are hardly distinguishable, indicating a nearly reversible elastic behavior. From fits with straight lines (see green dots) we deduce a Poisson's ratio of $-0.76\pm 0.02$. The corresponding numerically calculated strains for a finite sample with $2 \times 2 \times 2$ unit cells are shown as black dots. Using the identical structure parameters but assuming an infinite crystal (see previous section), we get a Poisson's ratio of $\nu=-0.79$. We note in passing that we have simultaneously measured the axial force from a load cell (not depicted). Comparing with theory, we obtain a metamaterial Young's modulus of $0.015\%\, E$ with the above bulk Young's modulus $E$. 

\section{Microscopic dilational metamaterials}

\begin{figure}[h]
\begin{center}
\includegraphics[scale=0.8]{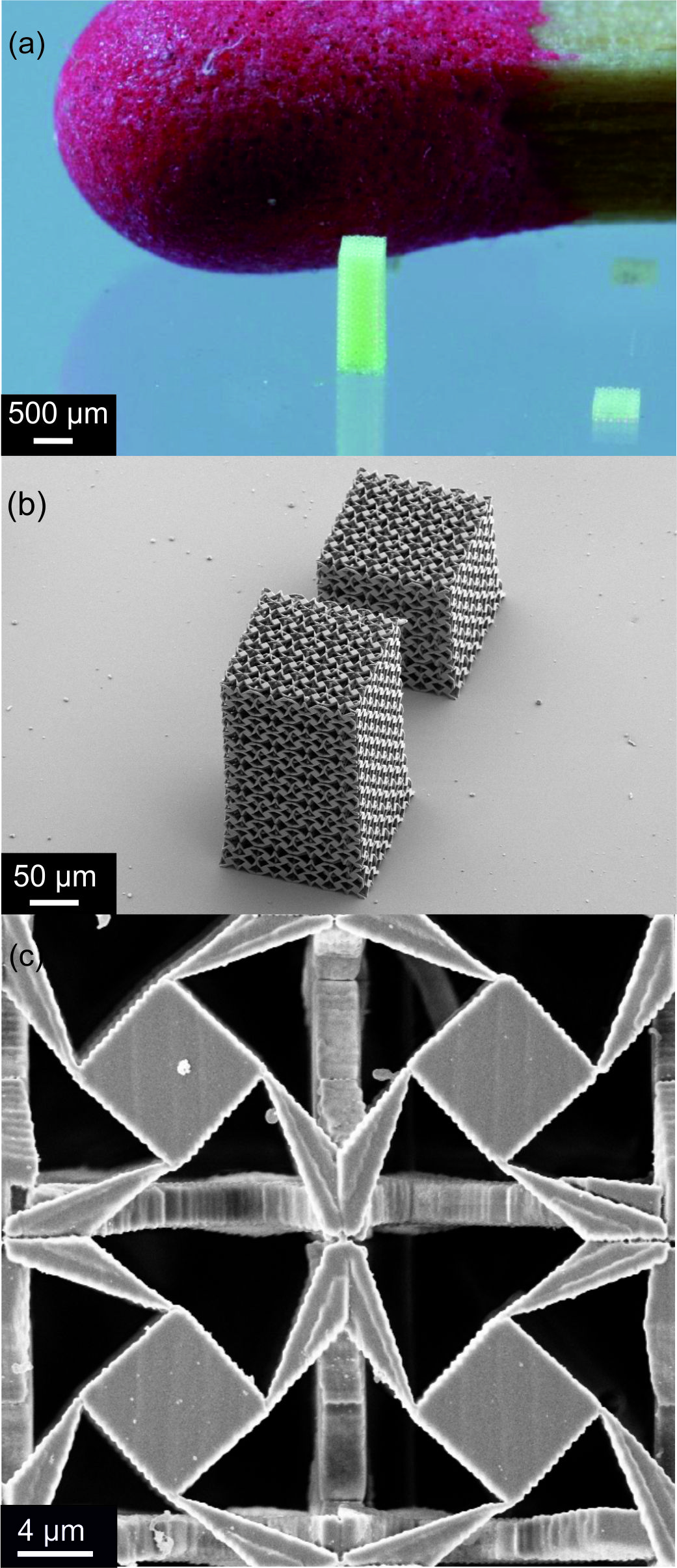}

\caption{Gallery of polymer dilational metamaterial microstructures with different sizes and aspect ratios following the blueprint illustrated in Fig.\,1 (without the small cubic tracking markers), all fabricated by 3D dip-in direct laser writing. (a) Photograph of a structure with $3\times3\times9$ unit cells (and a smaller one on the right-hand side) $a=180\, \rm{\upmu m}$. (b) Electron micrograph of two microstructure samples with overall aspect ratios of 1:1 and 2:1, respectively $a=35 \,\rm{\upmu m}$. (c) Magnified view onto one unit cell of the structure, revealing details within the unit cell (compare Fig.\,\ref{Fig1}), $a=35\, \rm{\upmu m}$.}
\label{Fig8}
\end{center}
\end{figure} 

The structures shown in the preceding section have validated our theoretical blueprint of a three-dimensional dilational metamaterial but they hardly qualify as a ``material'' in the normal sense. Thus, it is interesting to ask whether corresponding structures with lattice constants $a$ that are two to three orders of magnitude smaller are in reach. Also, it would be highly desirable to obtain structures containing a larger total number of unit cells. We have thus also fabricated microscopic structures based on the same blueprint (without the small blocks for tracking).
For fabricating such microscopic dilational metamaterial samples, photoresist samples are prepared by drop-casting the commercially available negative-tone photoresist ``IP-Dip'' (Nanoscribe GmbH, Germany) on diced silicon wafers ($22\, {\rm mm} \times 22\,\rm mm$). We use the commercial direct laser writing (DLW) system Photonic Professional GT (Nanoscribe GmbH, Germany). In this instrument, the liquid photoresist is polymerized via two-photon absorption using a $40\,\rm MHz$ frequency-doubled Erbium fiber laser with a pulse duration of $90\,\rm fs$. To avoid depth-dependent aberrations, the objective lens (with numerical aperture $\rm NA = 1.3$ or $\rm NA = 0.8$, Carl Zeiss) is directly dipped into the resist. The laser focus is scanned using a set of pivoted galvo mirrors. Structural data are again created in STL file format using the open-source software Blender and COMSOL Multiphysics. Due to the demanding critical distances of the mechanical metamaterials, the scan raster is set to $200\,\rm nm$ ($400\,\rm nm$) laterally and $300\,\rm nm$ ($800\,\rm nm$) axially for the $\rm NA = 1.3$ ($\rm NA = 0.8$) objective lens. Each individual layer is scanned in the so-called skywriting mode, i.e., while the laser focus is scanned continuously, the laser power is switched between $0\,\rm mW$ (no exposure) and about $13\,\rm mW$ or higher (exposure) to build up the fine features of the metamaterial. The writing speed is set to $20\,\rm mm/s$. After DLW of the preprogrammed pattern, the exposed sample is developed for 20 minutes in isopropanol and acetone. The process is finished in a supercritical point dryer to avoid capillary forces during drying.

Optical and electron micrographs of different samples are depicted in Fig.\,\ref{Fig8}. The sample in panel (a) has overall dimensions of $0.54\,{\rm mm}\times 0.54\,{\rm  mm}\times 1.62\,{\rm mm}$ ($3\times3\times9$ unit cells) yet, at the same time, minimum feature sizes in the sub-micron range. 
Due to the smaller lattice constants, it is not possible though to resolve the details within the unit cell to track the positions of the small marker cubes for measuring the Poisson's ratio as done for the macroscopic samples. 

\section{Conclusion}
We have designed, fabricated, and characterized a three-dimensional microstructure based on a simple-cubic translational lattice that effectively acts as an auxetic, converging for small internal connections $d/a\rightarrow 0$ to the ultimate limit of an isotropic three-dimensional dilational metamaterial of $\nu=-1$. Our experiments approach this limit. Interestingly, the Poisson's ratio becomes isotropic in the limit $d/a\rightarrow 0$, whereas the acoustic phase velocity and other elastic properties remain anisotropic.   
If fabricated in larger volumes and composed of different constituent materials, such dilational metamaterials might find applications in terms of shock absorbers.
In our treatment, we have derived all elements of the effective metamaterial elasticity tensor and hence all elastic parameters by comparing the phonon band structure in the long-wavelength limit with continuum mechanics of homogeneous media. This parameter retrieval could be of interest for other cubic-symmetry elastic metamaterials beyond the specific example discussed here.
 
\section{Acknowledgements}
We thank the DFG-Center for Functional Nanostructures (CFN), the Karlsruhe School of Optics $\&$ Photonics (KSOP) and the National Science foundation through grant DMS-1211359 for support.

\section*{References}

\end{document}